\documentclass[12pt]{article}

\baselineskip=\normalbaselineskip

\usepackage{mathrsfs,amsbsy,amssymb,latexsym,amsfonts,amsmath,amsthm}
\usepackage[nosort]{cite}
\usepackage{hyperref}

\usepackage{color}
\usepackage{ulem}

\usepackage{bm}
\usepackage{graphicx} %new
\allowdisplaybreaks[1]

\makeatletter
\catcode`\@=11
\@addtoreset{equation}{section}

\def\@seccntformat#1{\csname the#1\endcsname.~~}
\makeatother

\begin{document}
% !!!!!
\normalem
% !!!!!

\begin{titlepage}
  \renewcommand{\thefootnote}{\fnsymbol{footnote}}
   \begin{flushright}
     KEK-TH-2463, RBRC-1339
   \end{flushright}
  \vspace*{1.0cm}

\begin{center}
  {\Large \bf
    Energy of boundary of spacetime
  }
  \vspace{1.0cm}

\centerline{
{Yu Hamada${}^1$}%
\footnote{
  E-mail address: yuhamada@post.kek.jp
}
and
{Nobuyuki Matsumoto${}^2$}%
% {Nobuyuki Matsumoto}%
\footnote{
  E-mail address: nobuyuki.matsumoto@riken.jp} 
}

  \vskip 0.8cm
    ${}^1${\it Theory Center, High Energy Accelerator Research Organization (KEK),\newline Tsukuba 305-0801, Japan}
\vskip 0.1cm
  ${}^2${\it RIKEN/BNL Research center, Brookhaven National Laboratory,
    Upton, NY 11973, USA}
  \vskip 1.2cm

  \end{center}

  %%%%%%%%%%%%%%%%%%%%%%%%%%%%%%%%%%%%%%% 
  \begin{abstract}
    %%%%%%%%%%%%%%%%%%%%%%%%%%%%%%%%%%%%%%%
  We consider how 
  the energy 
  can be stored in the boundary of spacetime, 
  in particular 
  in a spherical bubble that can be made 
  by a quantum gravitational process.
  Our calculation is performed
  within the framework of 
  classical Einstein gravity
  by identifying the 
  Gibbons-Hawking-York term
  as the membrane action.
  We show that the energy of
  the bubble
  can be given consistently with the Schwarzschild metric.
  The solution of the consistency condition
  suggests positive
  membrane tension, 
  which explains why we do not 
  observe
  such topological defects
  in ordinary experiences
  and also gives a mechanism for 
  suppressing the spacetime with boundary
  in a dynamical way.
    %%%%%%%%%%%%%%%%%%%%%%%%%%%%%%%%%%%%%%% 
  \end{abstract}
  %%%%%%%%%%%%%%%%%%%%%%%%%%%%%%%%%%%%%%% 
\end{titlepage}

\pagestyle{empty}
\pagestyle{plain}

% \tableofcontents
\setcounter{footnote}{0}

% -----
\section{Introduction}
\label{sec:intro}
% ------
%

It is widely believed that quantum gravity allows change of topology
of spacetime \cite{Wheeler:1964qna, Geroch:1967fs,
Hawking:1978pog, 
Hawking:1979hw, Hawking:1979pi, 
Hawking:1984hk,
Hawking:1984pe, 
Sorkin:1985bh,
Hawking:1987mz,
Giddings:1987cg, 
Hawking:1988ae,
Coleman:1988tj,
Witten:1993yc,
DiFrancesco:1993cyw,
Aspinwall:1993yb,
Aspinwall:1993nu,
Giveon:1993ph,
Kiritsis:1994np,
Baez:1997zt,
Loll:2003yu}. 
However, we do not observe 
such topological 
objects in 
ordinary experiences,
and thus we can expect that 
such topology change only occurs at 
the size of the Planck length \cite{Hawking:1978pog,  Hawking:1979hw, Hawking:1979pi,
Hawking:1987mz,
Giddings:1987cg}
and the objects are difficult to 
exist in larger size.
It is notable that
most studies in the literature have focused on topology changes without the appearance of 
the boundary, such as wormholes and baby universes.
As will be referred to in discussion,
one of the reasons to avoid configurations
with boundary should be the geodesic incompleteness.
However, such restriction
in the summation over all 
configurations of spacetime
disagrees with the characteristic of quantum mechanics
that we allow
all the possible configurations
in principle.
When we take such a viewpoint, 
it is desirable
to have a mechanism to suppress
the configurations with boundary
at large scale
by the dynamics of the gravity.

In this letter,
we explore such a mechanism
by hypothesizing that
the boundaries,
which we observe as 
the defects of the spacetime,
carry energy
\cite{Vilenkin:1981zs,Gott:1984ef,Hiscock:1985uc,Linet:1985mvm,Hindmarsh:1994re,LaHaye:2020lsb},
and it costs larger energy for larger size.
We show that 
this picture can be made precise
within classical Einstein gravity
by a framework 
to calculate the energy of the defect.
Our calculation
is based on the two observations:
(i) Once the topological defect is settled 
in the form that can be
described by classical gravity,
we assign the Gibbons-Hawking-York term 
\cite{York:1972sj,Gibbons:1976ue}
to the added boundary.
(ii) The Gibbons-Hawking-York term can be 
identified as the Nambu-Goto action
\cite{Nambu:1977, Goto:1971ce}
describing a membrane
whose tension is given by the extrinsic curvature.
This identification then allows us to analyze the boundary as a dynamical membrane \cite{Jourjine:1983du} 
and calculate its energy 
from the membrane tension.

As an example,
we consider a spherically 
symmetric bubble
whose interior is 
outside the spacetime 
and thus absolute nothing \cite{Witten:1981gj}.
Our picture is however different
from conventional pictures of the 
bubble of nothing
as we consider its boundary 
and interpret 
boundary term as the membrane action,
from which we calculate the energy of the defect.
It should also be noted that
we work in four dimension.
At this stage, it is mainly for simplicity
to consider this particular shape of the defect.
Nevertheless, since the symmetry greatly simplifies the 
argument, we believe that the spherical defect gives a good testing ground to check
the validity of the identification.

Below, we show that the energy of 
the bubble can be given 
consistently with the spacetime 
described by the Schwarzschild metric.
The consistency condition 
equates the energy of the bubble 
to the Schwarzschild mass,
which has a unique solution
suggesting that
the surface of the boundary be located outside the event horizon. 
This means that they are
distinguishable from
microscopic black holes 
\cite{Bowick:1988xh, Coleman:1991sj, Adler:2001vs}.
Furthermore,
the solution shows the tension is positive,
which indicates that 
our universe is stable against the creation of such bubbles
as it tends to shrink,
not grow and swallow the whole universe
\cite{Witten:1981gj,
Aharony:2002cx,Birmingham:2002st,Balasubramanian:2002am,Horowitz:2002cx,Dine:2004uw,Jones:2007uj,He:2007ji,Aman:2008te,Rubakov:2012tj,Blanco-Pillado:2016xvf,Draper:2021qtc,Draper:2021ujg,Bomans:2021ara}.
The radius of the defect will be proportional to 
the stored energy, 
and therefore 
when the bubble is created
by quantum fluctuation of the order of the Planck mass,
its size will be the order of the Planck length.
Then, 
since the bubbles tend to shrink, 
they are always smaller than the order of the Planck length.
This picture thus gives a possible explanation 
why we do not observe such topological defects in ordinary experiences.

%%%%
% ------
\section{Energy of the spherically symmetric bubble}
\label{sec:energy}
% ------
%
We consider an 
appearance of a boundary
when spacetime undergoes a topology change.
In the framework of the Einstein gravity,
we assign the Gibbons-Hawking-York term
to the spacetime boundary:
\begin{align}
  S_B \equiv  
  {1\over 8\pi}\int_{\partial V} d^3 y \,
  \sqrt{\vert {\rm det}\, h_{ab} \vert } \ \varepsilon K.
  \label{eq:gh_term}
\end{align}
Here $h_{ab}$ is the induced metric on the boundary $\partial V$:
\begin{align}
  ds^2|_{\partial V} = h_{ab}dy^a dy^b,
\end{align}
and $K$ is the trace of the extrinsic curvature, which can be written with the
unit outward normal $n_\mu$:
\begin{align}
  K = {n^\mu}_{;\mu}.
  \label{eq:ext}
\end{align}
The sign factor is defined as
\begin{align}
  \varepsilon \equiv n^2 = \left\{ \begin{array}{cl}
                                   -1 &\quad (n: {\rm timelike})\\
                                     1 &\quad (n: {\rm spacelike})
                                 \end{array}
  \right.
  .
\end{align}

The main proposal of this paper is 
to interpret the boundary action~\eqref{eq:gh_term}
as a membrane action with the
position-dependent tension 
$T\equiv-{\varepsilon K/(8\pi)}$.
Writing the embedding function of 
the membrane as $X^\mu(y)$, 
which is nothing but the location of the boundary,
we have the induced metric:
\begin{align}
    h_{ab} = g_{\mu\nu} 
    \frac{dX^\mu}{dy^a}
    \frac{dX^\nu}{dy^b},
\end{align}
where $g_{\mu\nu}$ is the bulk spacetime metric.
Here $K$ can be considered as a functional of 
both $X^\mu$ and $g_{\mu\nu}$.
The Gibbons-Hawking-York term
then obtains the roles of both
the membrane action
and the surface term for the Einstein-Hilbert action.

\begin{figure}
    \centering
    \includegraphics[width=0.5\textwidth]{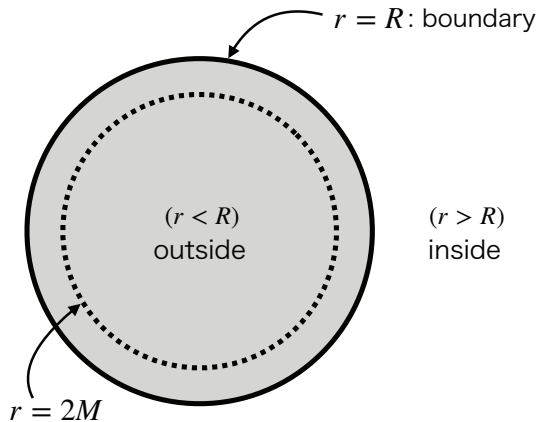}
    \caption{Schematic picture of a 
    spherically symmetric bubble
    in the spacetime at a time slice $t$.
    The boundary lies at $r=R$. 
    The region $r<R$ 
    (interior of the bubble, shaded in the figure) 
    is the {\it{outside}} of the spacetime.
    The solution of the consistency condition
    suggests $R$ be larger than
    the Schwarzschild radius $2M$.
    }
    \label{fig:bubble}
\end{figure}

To see that this identification 
gives a reasonable picture,
we consider the case that the
induced topological defect 
is spherically symmetric (Fig.~\ref{fig:bubble}).
To simplify the situation further, 
we consider a thought experiment
in which we make the spherical bubble
kept fixed
with no ordinary matter present. 
The spacetime should then be
static and spherically symmetric 
that is described by the Schwarzschild metric:
\begin{align}
  ds^2 = -\Big(1-\frac{2M}{r}\Big)dt^2 + \Big(1-\frac{2M}{r}\Big)^{-1} dr^2 + r^2 d\Omega^2,
  \label{eq:schw}
\end{align}
where $d\Omega$ is the line element of the unit 2-sphere.
In the ordinary description, 
we identify $M$ as the total energy of 
the matter sourcing gravity
\cite{Arnowitt:1959ah, Komar:1963svp}.
However, in our case, 
there is only the boundary of spacetime 
in the vacuum,
and thus we identify the boundary,
which we assume is a membrane,
to be the 
source of gravity.

In
the spherically symmetric case, 
the correspondence 
between the Gibbons-Hawking-York term
and the membrane action
is manifest because
$K$, and thus $T$, is constant 
over the surface due to 
the 
symmetry.
Since the spacetime 
is static with the Killing vector $\partial_t$
where $t$ corresponds to the 
Minkowski time 
in the asymptotically flat region,
we define the potential energy $E_{\rm pot}$ 
with respect to $t$.
Using the fact that the kinetic term is zero
since the membrane is held still,
we have
\begin{align}
  S_B 
  = - \int dt 
  \, E_{\rm pot} 
  = 
  {1\over 8\pi}\int dt 
  \sqrt{\vert g_{tt} \vert}\,
  \varepsilon K \Sigma,
  \label{eq:energy_extraction}
\end{align}
where
$g_{tt}\equiv -(1-2M/r)$ and
$\Sigma$ is the two-dimensional area of the membrane.
We do not add a factor
of $\sqrt{\vert g_{tt} \vert}$
in the left hand side
as in the case of the 
gravitational Hamiltonian \cite{Arnowitt:1962hi,Hawking:1995fd}.
Again since the spacetime is static,
we arrive at the following relation:
\begin{align}
  E_{\rm pot} = - 
  {1\over 8\pi}\sqrt{\vert g_{tt}\vert}   \, \varepsilon K \Sigma 
  = \sqrt{\vert g_{tt} \vert}\, T\Sigma,
  \label{eq:energy_relation}
\end{align}
which is the desired relation
giving the energy stored in the boundary of the spacetime.

%%%%%%%%%%%%%%%%%%%%%%
%%%%%%%%%%%%%%%%%%%%%%

% ------
\section{Consistency condition and its solution}
\label{sec:spherical}
% ------
%
We denote the location of the bubble as $r=R$.
The outward normal of the bubble is
\begin{align}
  n_\mu = \frac{- \varepsilon \, \partial_\mu r}{\sqrt{\vert1-2M/r\vert}} \Bigg\vert_{r=R}, 
  \label{eq:outward}
\end{align}
where $\varepsilon = {\rm sign}(1-2M/R)$.
The overall sign of $n_\mu$ is defined in such a way that
\begin{align}
  n^\mu \partial_\mu r = -\sqrt{\vert1-2M/R\vert} < 0,
\end{align}
reflecting the fact that $r=0$ is the {\it outside} when seen from the spacetime.
From eq.~\eqref{eq:ext}, we have
\begin{align}
  K = 
  -\varepsilon \cdot \frac{1}{R}
  \frac{1}{\sqrt{\vert 1-2M/R\vert}}
  \Big( 2-\frac{3M}{R} \Big),
\end{align}
which gives the tension:
\begin{align}
  T = \frac{1}{8\pi R}
  \frac{1}{\sqrt{\vert 1-2M/R\vert}}
  \Big( 2-\frac{3M}{R} \Big).
 \label{eq:string_tension_sol}
\end{align}
Note that the tension $T$  is positive for $R>3M/2$.

We now have the two explicit expressions
for the energy sourcing gravity.
One is simply the mass parameter $M$
in the Schwarzschild metric~\eqref{eq:schw},
which is identical to the ADM mass \cite{Arnowitt:1959ah} for this spacetime.
The other is the internal energy of the membrane:
\begin{align}
  E_{\rm pot} = \frac{R}{2}\Big(2-\frac{3M}{R}\Big),
\end{align}
where we set $\Sigma=4\pi R^2$,
which is obtained by analyzing the boundary action.
For consistency, these two quantities should be equal:
\begin{align}
  M = \frac{R}{2}\Big(2-\frac{3M}{R}\Big),
  \label{eq:consistency}
\end{align}
from which we obtain
the relation between
the radius and the total mass of the bubble:
\begin{align}
  R=5M/2.
\end{align}
This unique solution 
gives us important consequences:
Firstly, 
the spherical defect is 
not covered by the horizon, $R>2M$,
and thus it is distinguishable
from microscopic black holes.
This is important because it makes us possible
to directly interact with the remnant
of a quantum gravitational effect.
Secondly, the tension is positive,
and thus classically the bubble tends to shrink;
otherwise, our universe becomes unstable
as the bubble can grow until it swallows the whole universe,
which adds another consistency
to this picture.
We thus see that the identification of the Gibbons-Hawking-York term
with the membrane action gives
a consistent description of the energy 
stored in the boundary of spacetime.

% ------
\section{Discussion}
\label{sec:summary}
% ------
%
In this letter, we considered a framework to calculate the energy
of the boundary of spacetime 
and showed that the spherically symmetric bubble defect
can be described consistently with the Schwarzschild metric.
However, it is true that
many assumptions are made in the argument.
For example, as emphasized in the main text, we thoroughly used the properties of the Schwarzschild solution in identifying the potential energy, eq.~\eqref{eq:energy_extraction}. 
The notion of energy 
itself is a subtle topic
in general relativity
(see, e.g., 
\cite{lyndenbell:1985,Brown:1992br,Nikolic:2014kga,Aoki:2020prb,BeltranJimenez:2021kpj}) and 
we may need a more sophisticated formula to apply the argument for general cases.
Another assumption we implicitly made is 
that the cosmological constant
is irrelevant.
Generalization in this direction
is interesting because
its small value has been related
to the topological nature of 
quantum gravity \cite{Hawking:1984hk, Coleman:1988tj}. 

Another subtlety related to the argument is the
choice of boundary action.
Though we used the Gibbons-Hawking-York term
because it is the simplest boundary term
for the variational problem 
of the metric to be well-defined, 
the boundary cosmological constant
and boundary curvature terms may be added in principle
\cite{Gibbons:1976ue}.\footnote{
  We thank the referee of PTEP
  for raising this important point.
  See also \cite{Poisson:2009pwt}
  for a discussion on the boundary cosmological constant.
}
However, it is possible to make an argument
for the simplest choice.
Let us consider
a general ADM decomposition \cite{Arnowitt:1959ah}
for a timelike coordinate $t$:
\begin{align}
  ds^2 = - N^2 dt^2 + h_{ab}(N^a dt + dy^a) (N^b dt + dy^b).
\end{align}
It is known that \cite{York:1972sj}:
\begin{align}
  \sqrt{-g} R = N\sqrt{h} ( ^3R + K_{ab}K^{ab} -K^2 )
  + 2\partial_t (\sqrt{h} K) - 2 \partial_a \{\sqrt{h} (D^a N + N^a K) \},
  \label{eq:R_decomp}
\end{align}
where $^3 R$ is the Ricci scalar
on the $t$-constant surface $\Sigma_t$ and $K_{ab}$ the extrinsic curvature.
We now consider the spacetime bounded by
the two $t$-constant surfaces $\Sigma_{t_1}$, $\Sigma_{t_2}$.
The second term of eq.~\eqref{eq:R_decomp} suggests the
above-mentioned 
need for the Gibbons-Hawking-York term on 
$\Sigma_{t_1}$ and $\Sigma_{t_2}$
to make the variational problem well-defined.
On the other hand, if we add additional terms such as curvature or cosmological constant terms on the boundaries,
they cannot be canceled by the total derivatives in the bulk.
The situation remains the same after moving to the
Hamiltonian formalism,
in which one may interpret the 
additional terms as
an instant pulse potential 
for the induced metric
at the initial and final
surfaces.
Then, in this particular problem,
the potential depends on
the artificial times slices.
We consider such a theory to be 
an unusual extension from the conventional mechanics.

Allowing the boundary degrees of freedom to exist,
there are many topics that can be explored.
An interesting question
is whether such bubbles 
can be stabilized by 
a quantum effect at the Planck scale
because if such stable bubbles exist,
they can be a candidate of 
dark matter.
To see that this is a viable possibility, 
let us assume that
the bubbles were created and stabilized 
by some quantum gravitational process
before the inflationary epoch.
Since the relevant scale is the Planck scale $M_P$, the number density should be $n_b\sim M_P^3$ (before the inflation)
and the stabilized mass $M\sim M_P$.
During the inflationary expansion, the number density is exponentially diluted by the factor $\exp{(-3 N_e)}$ with $N_e$ the number of 
e-foldings.
After the inflation ends and the universe is reheated, the yield variable
is given by
$Y_b\equiv n_b/s \sim M_P^3 \exp{(-3 N_e)}/ (g_{S} T_{RH}^3)$, 
where $T_{RH}$ is the reheating temperature and $g_{S}(\simeq \mathcal{O}(10^2))$ the effective number of relativistic degrees of freedom.
The fraction of the energy density of the bubble to the critical density
at the present universe then becomes
\begin{align}
    \Omega_{b}h^2 \sim Y_{b} \,
\frac{M_P}{\mathrm{GeV}}\times 2.8\times 10^8 %\nonumber \\
 \sim 0.1 \bigg(\frac{10^6 \, \mathrm{GeV}}{T_{RH}} \bigg)^3  \exp(151 -3N_e) ,\label{eq:relic}
\end{align}
where $h$ is the dimensionless Hubble parameter.
By identifying eq.~\eqref{eq:relic} with the observed relic abundance of dark matter: $\Omega_{DM}h^2 \simeq 0.12$~\cite{Planck:2018vyg} and taking the e-foldings $N_e \simeq 50$ \cite{Planck:2015sxf},
we are led to the prediction $T_{RH} \sim 10^6 \, \mathrm{GeV}$,
whose value is compatible with leptogenesis scenarios with low reheating temperatures
(see, e.g., 
\cite{Davidson:2008bu} for a review).

It is also important to consider the interaction of the bubbles with ordinary matters \cite{Jourjine:1983du}
for detection.
In this regard,
the bubbles may be thought of as particles.
Note that
geodesics around the bubble 
cannot be extended into the 
$r<R$ region,
and in this sense the spacetime is singular.
For this, it should be crucial 
to specify an appropriate 
boundary condition to determine 
the behavior of the
incoming particles (or the fields)
at the surface of the bubbles
so that we can obtain predictions
for the entire physical processes.
In such investigations,
the full dynamical treatment of the bubble
should also be important,
in which we 
simultaneously solve the equations of motion of the membrane (i.e., of the embedding functions $X^\mu(y)$)
and the time-dependent Einstein equation.

Finally, since the bubble creation assumes
a non-perturbative dynamics of quantum gravity,
it is intriguing in turn to consider what kind of theory
can provide this process microscopically,
and to confirm that the bubble can be described
consistently from its creation to annihilation.
Non-perturbative dynamics of spacetime
has been studied extensively in two-dimensional gravity
with matrix models,
in which the correspondence between the spacetime boundaries
and $D$-branes (or $D$-instantons) is pointed out \cite{Hanada:2004im}.
We consider establishing our picture
with such frameworks to be a promising next step
towards probing quantum gravity.

Studies along these lines are in progress and will be reported elsewhere.

% ------
\section*{Acknowledgments}
% ------
%
The authors thank  
Yuta Hamada and Taku Izubuchi
for valuable discussions.
NM is also grateful to the radiation laboratory of 
RIKEN for their hospitality during the initial stage of this work.
The authors further thank the referee of PTEP
for valuable comments that helped improve the manuscript.
YH is supported by
JSPS KAKENHI Grant Number JP21J0111, 
and NM is supported by JP22H01222 and the Special Postdoctoral Researchers Program of RIKEN.
%

% \appendix

%%%%%%%%%%%%%%%%%%%%%%%%%%%%%%%%%%%%%%% 
\baselineskip=0.9\normalbaselineskip
%%%%%%%%%%%%%%%%%%%%%%%%%%%%%%%%%%%%%%%

%%%%%%%%%%%%%%%%%%%%%%%%%%%%%%%%%%%
% \bibliographystyle{utphys}
% \bibliography{ref}

\begin{thebibliography}{99}
  \setlength{\itemsep}{-2pt}
  %%%%%%%%%%%%%%%%%%%%%%%%%%%%%%%%%%%%%%%
  %%%%%%%%%%%%%%%%%%%%%%%%%%%%%%%%%%%%%%%


%\cite{Wheeler:1964qna}
\bibitem{Wheeler:1964qna}
J.~A.~Wheeler,
``Geometrodynamics and the issue of final state,''
in Relativity, Groups and Topology (ed. by B. S. DeWitt and C. M. DeWitt), Gordon and Breach, New York,
328-504 (1964).
%5 citations counted in INSPIRE as of 05 Dec 2022

%\cite{Geroch:1967fs}
\bibitem{Geroch:1967fs}
R.~P.~Geroch,
``Topology in general relativity,''
J. Math. Phys. \textbf{8}, 782-786 (1967).
% doi:10.1063/1.1705276
%247 citations counted in INSPIRE as of 05 Dec 2022

%\cite{Hawking:1978pog}
\bibitem{Hawking:1978pog}
S.~W.~Hawking,
``Space-Time Foam,''
Nucl. Phys. B \textbf{144}, 349-362 (1978).
% doi:10.1016/0550-3213(78)90375-9
%391 citations counted in INSPIRE as of 01 Sep 2022

%\cite{Hawking:1979hw}
\bibitem{Hawking:1979hw}
S.~W.~Hawking, D.~N.~Page and C.~N.~Pope,
``The propagation of particles in space-time foam,''
Phys. Lett. B \textbf{86}, 175-178 (1979).

%\cite{Hawking:1979pi}
\bibitem{Hawking:1979pi}
S.~W.~Hawking, D.~N.~Page and C.~N.~Pope,
``Quantum Gravitational Bubbles,''
Nucl. Phys. B \textbf{170}, 283-306 (1980).
% doi:10.1016/0550-3213(80)90151-0
%148 citations counted in INSPIRE as of 01 Sep 2022

%\cite{Hawking:1984hk}
\bibitem{Hawking:1984hk}
S.~W.~Hawking,
``The Cosmological Constant Is Probably Zero,''
Phys. Lett. B \textbf{134}, 403 (1984).
% doi:10.1016/0370-2693(84)91370-4
%532 citations counted in INSPIRE as of 05 Dec 2022

%\cite{Hawking:1984pe}
\bibitem{Hawking:1984pe}
S.~W.~Hawking,
%``Nontrivial Topologies in Quantum Gravity,''
Nucl. Phys. B \textbf{244}, 135-146 (1984).
doi:10.1016/0550-3213(84)90185-8
%76 citations counted in INSPIRE as of 01 Sep 2022

%\cite{Sorkin:1985bh}
\bibitem{Sorkin:1985bh}
R.~D.~Sorkin,
``On Topology Change and Monopole Creation,''
Phys. Rev. D \textbf{33}, 978-982 (1986).
% doi:10.1103/PhysRevD.33.978
%55 citations counted in INSPIRE as of 05 Dec 2022

%\cite{Hawking:1987mz}
\bibitem{Hawking:1987mz}
S.~W.~Hawking,
``Quantum Coherence Down the Wormhole,''
Phys. Lett. B \textbf{195}, 337 (1987).
% doi:10.1016/0370-2693(87)90028-1
%301 citations counted in INSPIRE as of 05 Dec 2022

%\cite{Giddings:1987cg}
\bibitem{Giddings:1987cg}
S.~B.~Giddings and A.~Strominger,
``Axion Induced Topology Change in Quantum Gravity and String Theory,''
Nucl. Phys. B \textbf{306}, 890-907 (1988).
% doi:10.1016/0550-3213(88)90446-4
%545 citations counted in INSPIRE as of 05 Dec 2022

%\cite{Hawking:1988ae}
\bibitem{Hawking:1988ae}
S.~W.~Hawking,
``Wormholes in Space-Time,''
Phys. Rev. D \textbf{37}, 904-910 (1988).
% doi:10.1103/PhysRevD.37.904
%461 citations counted in INSPIRE as of 05 Dec 2022

 % \cite{Coleman:1988tj}
\bibitem{Coleman:1988tj}
S.~R.~Coleman,
``Why There Is Nothing Rather Than Something: A Theory of the Cosmological Constant,''
Nucl. Phys. B \textbf{310}, 643-668 (1988).
% doi:10.1016/0550-3213(88)90097-1
%885 citations counted in INSPIRE as of 31 Aug 2022

%\cite{Witten:1993yc}
\bibitem{Witten:1993yc}
E.~Witten,
``Phases of N=2 theories in two-dimensions,''
Nucl. Phys. B \textbf{403}, 159-222 (1993)
% doi:10.1016/0550-3213(93)90033-L
[arXiv:hep-th/9301042 [hep-th]].
%1380 citations counted in INSPIRE as of 05 Dec 2022

%\cite{DiFrancesco:1993cyw}
\bibitem{DiFrancesco:1993cyw}
P.~Di Francesco, P.~H.~Ginsparg and J.~Zinn-Justin,
``2-D Gravity and random matrices,''
Phys. Rept. \textbf{254}, 1-133 (1995)
% doi:10.1016/0370-1573(94)00084-G
[arXiv:hep-th/9306153 [hep-th]].
%899 citations counted in INSPIRE as of 05 Dec 2022

%\cite{Aspinwall:1993yb}
\bibitem{Aspinwall:1993yb}
P.~S.~Aspinwall, B.~R.~Greene and D.~R.~Morrison,
``Multiple mirror manifolds and topology change in string theory,''
Phys. Lett. B \textbf{303}, 249-259 (1993)
% doi:10.1016/0370-2693(93)91428-P
[arXiv:hep-th/9301043 [hep-th]].
%122 citations counted in INSPIRE as of 05 Dec 2022

%\cite{Aspinwall:1993nu}
\bibitem{Aspinwall:1993nu}
P.~S.~Aspinwall, B.~R.~Greene and D.~R.~Morrison,
``Calabi-Yau moduli space, mirror manifolds and space-time topology change in string theory,''
Nucl. Phys. B \textbf{416}, 414-480 (1994)
% doi:10.1016/0550-3213(94)90321-2
[arXiv:hep-th/9309097 [hep-th]].
%306 citations counted in INSPIRE as of 05 Dec 2022

%\cite{Giveon:1993ph}
\bibitem{Giveon:1993ph}
A.~Giveon and E.~Kiritsis,
``Axial vector duality as a gauge symmetry and topology change in string theory,''
Nucl. Phys. B \textbf{411}, 487-508 (1994)
% doi:10.1016/0550-3213(94)90460-X
[arXiv:hep-th/9303016 [hep-th]].
%133 citations counted in INSPIRE as of 05 Dec 2022

%\cite{Kiritsis:1994np}
\bibitem{Kiritsis:1994np}
E.~Kiritsis and C.~Kounnas,
``Dynamical topology change in string theory,''
Phys. Lett. B \textbf{331}, 51-62 (1994)
% doi:10.1016/0370-2693(94)90942-3
[arXiv:hep-th/9404092 [hep-th]].
%93 citations counted in INSPIRE as of 05 Dec 2022

%\cite{Baez:1997zt}
\bibitem{Baez:1997zt}
J.~C.~Baez,
``Spin foam models,''
Class. Quant. Grav. \textbf{15}, 1827-1858 (1998)
% doi:10.1088/0264-9381/15/7/004
[arXiv:gr-qc/9709052 [gr-qc]].
%353 citations counted in INSPIRE as of 05 Dec 2022

%\cite{Loll:2003yu}
\bibitem{Loll:2003yu}
R.~Loll and W.~Westra,
``Space-time foam in 2-D and the sum over topologies,''
Acta Phys. Polon. B \textbf{34}, 4997-5008 (2003)
[arXiv:hep-th/0309012 [hep-th]].
%15 citations counted in INSPIRE as of 05 Dec 2022

%\cite{Vilenkin:1981zs}
\bibitem{Vilenkin:1981zs}
A.~Vilenkin,
``Gravitational Field of Vacuum Domain Walls and Strings,''
Phys. Rev. D \textbf{23}, 852-857 (1981).
% doi:10.1103/PhysRevD.23.852
%1053 citations counted in INSPIRE as of 05 Dec 2022

%\cite{Gott:1984ef}
\bibitem{Gott:1984ef}
J.~R.~Gott, III,
``Gravitational lensing effects of vacuum strings: Exact solutions,''
Astrophys. J. \textbf{288}, 422-427 (1985).
% doi:10.1086/162808
%666 citations counted in INSPIRE as of 05 Dec 2022

%\cite{Hiscock:1985uc}
\bibitem{Hiscock:1985uc}
W.~A.~Hiscock,
``Exact Gravitational Field of a String,''
Phys. Rev. D \textbf{31}, 3288-3290 (1985).
% doi:10.1103/PhysRevD.31.3288
%432 citations counted in INSPIRE as of 05 Dec 2022

%\cite{Linet:1985mvm}
\bibitem{Linet:1985mvm}
B.~Linet,
``The Static Metrics with Cylindrical Symmetry Describing a Model of Cosmic Strings,''
Gen. Rel. Grav. \textbf{17}, 1109-1115 (1985).
% doi:10.1007/BF00774211
%251 citations counted in INSPIRE as of 05 Dec 2022

%\cite{Hindmarsh:1994re}
\bibitem{Hindmarsh:1994re}
M.~B.~Hindmarsh and T.~W.~B.~Kibble,
``Cosmic strings,''
Rept. Prog. Phys. \textbf{58}, 477-562 (1995)
% doi:10.1088/0034-4885/58/5/001
[arXiv:hep-ph/9411342 [hep-ph]].
%1046 citations counted in INSPIRE as of 05 Dec 2022

%\cite{LaHaye:2020lsb}
\bibitem{LaHaye:2020lsb}
M.~LaHaye and E.~Poisson,
``Self-force from a conical singularity, without renormalization,''
Phys. Rev. D \textbf{101}, no.10, 104047 (2020)
% doi:10.1103/PhysRevD.101.104047
[arXiv:2001.00430 [gr-qc]].
%7 citations counted in INSPIRE as of 05 Dec 2022

%\cite{York:1972sj}
\bibitem{York:1972sj}
J.~W.~York, Jr.,
``Role of conformal three geometry in the dynamics of gravitation,''
Phys. Rev. Lett. \textbf{28}, 1082-1085 (1972).
% doi:10.1103/PhysRevLett.28.1082
%1045 citations counted in INSPIRE as of 31 Aug 2022

%\cite{Gibbons:1976ue}
\bibitem{Gibbons:1976ue}
G.~W.~Gibbons and S.~W.~Hawking,
``Action Integrals and Partition Functions in Quantum Gravity,''
Phys. Rev. D \textbf{15}, 2752-2756 (1977).
% doi:10.1103/PhysRevD.15.2752
%2932 citations counted in INSPIRE as of 31 Aug 2022

%\cite{Nambu:1977}
\bibitem{Nambu:1977}
Y.~Nambu, 
``Duality and Hydrodynamics,''
Lectures at the Copenhagen conference (1970).

%\cite{Goto:1971ce}
\bibitem{Goto:1971ce}
T.~Goto,
``Relativistic quantum mechanics of one-dimensional mechanical continuum and subsidiary condition of dual resonance model,''
Prog. Theor. Phys. \textbf{46}, 1560-1569 (1971).
% doi:10.1143/PTP.46.1560
%673 citations counted in INSPIRE as of 31 Aug 2022

%\cite{Jourjine:1983du}
\bibitem{Jourjine:1983du}
A.~N.~Jourjine,
``On the coupling of matter and gravity to the boundary of space-time,''
Phys. Lett. B \textbf{136}, 237 (1984).
% doi:10.1016/0370-2693(84)91153-5
%4 citations counted in INSPIRE as of 05 Dec 2022

%\cite{Witten:1981gj}
\bibitem{Witten:1981gj}
E.~Witten,
``Instability of the Kaluza-Klein Vacuum,''
Nucl. Phys. B \textbf{195}, 481-492 (1982).
% doi:10.1016/0550-3213(82)90007-4
%504 citations counted in INSPIRE as of 02 Sep 2022

%\cite{Bowick:1988xh}
\bibitem{Bowick:1988xh}
M.~J.~Bowick, S.~B.~Giddings, J.~A.~Harvey, G.~T.~Horowitz and A.~Strominger,
``Axionic Black Holes and a Bohm-Aharonov Effect for Strings,''
Phys. Rev. Lett. \textbf{61}, 2823 (1988).
% doi:10.1103/PhysRevLett.61.2823
%196 citations counted in INSPIRE as of 05 Dec 2022

%\cite{Coleman:1991sj}
\bibitem{Coleman:1991sj}
S.~R.~Coleman, J.~Preskill and F.~Wilczek,
``Dynamical effect of quantum hair,''
Mod. Phys. Lett. A \textbf{6}, 1631-1642 (1991).
% doi:10.1142/S0217732391001767
%67 citations counted in INSPIRE as of 05 Dec 2022

%\cite{Adler:2001vs}
\bibitem{Adler:2001vs}
R.~J.~Adler, P.~Chen and D.~I.~Santiago,
``The Generalized uncertainty principle and black hole remnants,''
Gen. Rel. Grav. \textbf{33}, 2101-2108 (2001).
% doi:10.1023/A:1015281430411
% [arXiv:gr-qc/0106080 [gr-qc]].
%577 citations counted in INSPIRE as of 02 Sep 2022

%\cite{Aharony:2002cx}
\bibitem{Aharony:2002cx}
O.~Aharony, M.~Fabinger, G.~T.~Horowitz and E.~Silverstein,
``Clean time dependent string backgrounds from bubble baths,''
JHEP \textbf{07}, 007 (2002)
% doi:10.1088/1126-6708/2002/07/007
[arXiv:hep-th/0204158 [hep-th]].
%137 citations counted in INSPIRE as of 05 Dec 2022

%\cite{Birmingham:2002st}
\bibitem{Birmingham:2002st}
D.~Birmingham and M.~Rinaldi,
``Bubbles in anti-de Sitter space,''
Phys. Lett. B \textbf{544}, 316-320 (2002)
% doi:10.1016/S0370-2693(02)02261-X
[arXiv:hep-th/0205246 [hep-th]].
%62 citations counted in INSPIRE as of 05 Dec 2022

%\cite{Balasubramanian:2002am}
\bibitem{Balasubramanian:2002am}
V.~Balasubramanian and S.~F.~Ross,
``The Dual of nothing,''
Phys. Rev. D \textbf{66}, 086002 (2002)
% doi:10.1103/PhysRevD.66.086002
[arXiv:hep-th/0205290 [hep-th]].
%97 citations counted in INSPIRE as of 05 Dec 2022

%\cite{Horowitz:2002cx}
\bibitem{Horowitz:2002cx}
G.~T.~Horowitz and K.~Maeda,
``Colliding Kaluza-Klein bubbles,''
Class. Quant. Grav. \textbf{19}, 5543-5556 (2002)
% doi:10.1088/0264-9381/19/21/317
[arXiv:hep-th/0207270 [hep-th]].
%16 citations counted in INSPIRE as of 05 Dec 2022


%\cite{Dine:2004uw}
\bibitem{Dine:2004uw}
M.~Dine, P.~J.~Fox and E.~Gorbatov,
``Catastrophic decays of compactified space-times,''
JHEP \textbf{09}, 037 (2004)
% doi:10.1088/1126-6708/2004/09/037
[arXiv:hep-th/0405190 [hep-th]].
%17 citations counted in INSPIRE as of 05 Dec 2022

%\cite{Jones:2007uj}
\bibitem{Jones:2007uj}
G.~C.~Jones and J.~E.~Wang,
``How to stop (worrying and love) the bubble: Boundary changing solutions,''
JHEP \textbf{07}, 076 (2007)
% doi:10.1088/1126-6708/2007/07/076
[arXiv:hep-th/0701183 [hep-th]].
%2 citations counted in INSPIRE as of 05 Dec 2022

%\cite{He:2007ji}
\bibitem{He:2007ji}
J.~He and M.~Rozali,
``On bubbles of nothing in AdS/CFT,''
JHEP \textbf{09}, 089 (2007)
% doi:10.1088/1126-6708/2007/09/089
[arXiv:hep-th/0703220 [hep-th]].
%28 citations counted in INSPIRE as of 05 Dec 2022

%\cite{Aman:2008te}
\bibitem{Aman:2008te}
J.~E.~Aman, S.~Aminneborg, I.~Bengtsson and N.~Pidokrajt,
``Anti-de Sitter Quotients, Bubbles of Nothing, and Black Holes,''
Gen. Rel. Grav. \textbf{40}, 2557-2567 (2008)
% doi:10.1007/s10714-008-0639-z
[arXiv:0801.4214 [hep-th]].
%1 citations counted in INSPIRE as of 05 Dec 2022

%\cite{Rubakov:2012tj}
\bibitem{Rubakov:2012tj}
V.~A.~Rubakov and M.~Y.~Kuznetsov,
``Fermions and Kaluza-Klein vacuum decay: a toy model,''
Theor. Math. Phys. \textbf{175}, 489-498 (2013)
% doi:10.1007/s11232-013-0040-2
[arXiv:1205.5184 [hep-th]].
%2 citations counted in INSPIRE as of 05 Dec 2022

%\cite{Blanco-Pillado:2016xvf}
\bibitem{Blanco-Pillado:2016xvf}
J.~J.~Blanco-Pillado, B.~Shlaer, K.~Sousa and J.~Urrestilla,
``Bubbles of Nothing and Supersymmetric Compactifications,''
JCAP \textbf{10}, 002 (2016)
% doi:10.1088/1475-7516/2016/10/002
[arXiv:1606.03095 [hep-th]].
%7 citations counted in INSPIRE as of 05 Dec 2022

%\cite{Draper:2021qtc}
\bibitem{Draper:2021qtc}
P.~Draper, I.~Garcia Garcia and B.~Lillard,
``De Sitter decays to infinity,''
JHEP \textbf{12}, 154 (2021)
% doi:10.1007/JHEP12(2021)154
[arXiv:2105.10507 [hep-th]].
%4 citations counted in INSPIRE as of 05 Dec 2022

%\cite{Draper:2021ujg}
\bibitem{Draper:2021ujg}
P.~Draper, I.~G.~Garcia and B.~Lillard,
``Bubble of nothing decays of unstable theories,''
Phys. Rev. D \textbf{104}, no.12, 12 (2021)
% doi:10.1103/PhysRevD.104.L121701
[arXiv:2105.08068 [hep-th]].
%4 citations counted in INSPIRE as of 05 Dec 2022

%\cite{Bomans:2021ara}
\bibitem{Bomans:2021ara}
P.~Bomans, D.~Cassani, G.~Dibitetto and N.~Petri,
``Bubble instability of mIIA on $\mathrm{AdS}_4\times S^6$,''
SciPost Phys. \textbf{12}, no.3, 099 (2022)
% doi:10.21468/SciPostPhys.12.3.099
[arXiv:2110.08276 [hep-th]].
%14 citations counted in INSPIRE as of 05 Dec 2022

%\cite{Arnowitt:1959ah}
\bibitem{Arnowitt:1959ah}
R.~L.~Arnowitt, S.~Deser and C.~W.~Misner,
``Dynamical Structure and Definition of Energy in General Relativity,''
Phys. Rev. \textbf{116}, 1322-1330 (1959).
% doi:10.1103/PhysRev.116.1322
%680 citations counted in INSPIRE as of 05 Dec 2022

%\cite{Komar:1963svp}
\bibitem{Komar:1963svp}
A.~Komar,
``Positive-Definite Energy Density and Global Consequences for General Relativity,''
Phys. Rev. \textbf{129}, no.4, 1873 (1963).
% doi:10.1103/PhysRev.129.1873
%65 citations counted in INSPIRE as of 02 Sep 2022

%\cite{Arnowitt:1962hi}
\bibitem{Arnowitt:1962hi}
R.~L.~Arnowitt, S.~Deser and C.~W.~Misner,
``The Dynamics of general relativity,''
Gen. Rel. Grav. \textbf{40}, 1997-2027 (2008)
% doi:10.1007/s10714-008-0661-1
[arXiv:gr-qc/0405109 [gr-qc]].
%1868 citations counted in INSPIRE as of 05 Dec 2022

%\cite{Hawking:1995fd}
\bibitem{Hawking:1995fd}
S.~W.~Hawking and G.~T.~Horowitz,
``The Gravitational Hamiltonian, action, entropy and surface terms,''
Class. Quant. Grav. \textbf{13}, 1487-1498 (1996)
% doi:10.1088/0264-9381/13/6/017
[arXiv:gr-qc/9501014 [gr-qc]].
%548 citations counted in INSPIRE as of 05 Dec 2022

%\cite{lyndenbell:1985}
\bibitem{lyndenbell:1985}
D.~Lynden-Bell and J.~Katz,
``Gravitational field energy density for spheres and black holes,''
Monthly Notices of the Royal Astronomical Society \textbf{213},
no.1, 21-25 (1985).
% doi:10.1093/mnras/213.1.21P

%\cite{Brown:1992br}
\bibitem{Brown:1992br}
J.~D.~Brown and J.~W.~York, Jr.,
``Quasilocal energy and conserved charges derived from the gravitational action,''
Phys. Rev. D \textbf{47}, 1407-1419 (1993)
% doi:10.1103/PhysRevD.47.1407
[arXiv:gr-qc/9209012 [gr-qc]].
%1475 citations counted in INSPIRE as of 05 Dec 2022

%\cite{Nikolic:2014kga}
\bibitem{Nikolic:2014kga}
H.~Nikolic,
``The trivial solution of the gravitational energy-momentum tensor problem,''
[arXiv:1407.8028 [gr-qc]].
%3 citations counted in INSPIRE as of 05 Dec 2022

%\cite{Aoki:2020prb}
\bibitem{Aoki:2020prb}
S.~Aoki, T.~Onogi and S.~Yokoyama,
``Conserved charges in general relativity,''
Int. J. Mod. Phys. A \textbf{36}, no.10, 2150098 (2021)
% doi:10.1142/S0217751X21500986
[arXiv:2005.13233 [gr-qc]].
%19 citations counted in INSPIRE as of 05 Dec 2022

%\cite{BeltranJimenez:2021kpj}
\bibitem{BeltranJimenez:2021kpj}
J.~Beltr\'an Jim\'enez and T.~S.~Koivisto,
``Noether charges in the geometrical trinity of gravity,''
Phys. Rev. D \textbf{105}, no.2, L021502 (2022)
% doi:10.1103/PhysRevD.105.L021502
[arXiv:2111.04716 [gr-qc]].
%11 citations counted in INSPIRE as of 05 Dec 2022

%\cite{Poisson:2009pwt}
\bibitem{Poisson:2009pwt}
E.~Poisson,
``A Relativist's Toolkit: The Mathematics of Black-Hole Mechanics,''
Cambridge University Press, 2009.
% doi:10.1017/CBO9780511606601
%181 citations counted in INSPIRE as of 06 Feb 2023

%\cite{Planck:2018vyg}
\bibitem{Planck:2018vyg}
N.~Aghanim \textit{et al.} [Planck],
``Planck 2018 results. VI. Cosmological parameters,''
Astron. Astrophys. \textbf{641}, A6 (2020)
[erratum: Astron. Astrophys. \textbf{652}, C4 (2021)]
% doi:10.1051/0004-6361/201833910
[arXiv:1807.06209 [astro-ph.CO]].
%9227 citations counted in INSPIRE as of 05 Dec 2022

%\cite{Planck:2015sxf}
\bibitem{Planck:2015sxf}
P.~A.~R.~Ade \textit{et al.} [Planck],
``Planck 2015 results. XX. Constraints on inflation,''
Astron. Astrophys. \textbf{594}, A20 (2016)
% doi:10.1051/0004-6361/201525898
[arXiv:1502.02114 [astro-ph.CO]].
%2432 citations counted in INSPIRE as of 05 Dec 2022

%\cite{Davidson:2008bu}
\bibitem{Davidson:2008bu}
S.~Davidson, E.~Nardi and Y.~Nir,
``Leptogenesis,''
Phys. Rept. \textbf{466}, 105-177 (2008)
% doi:10.1016/j.physrep.2008.06.002
[arXiv:0802.2962 [hep-ph]].
%1028 citations counted in INSPIRE as of 05 Dec 2022


%\cite{Hanada:2004im}
\bibitem{Hanada:2004im}
M.~Hanada, M.~Hayakawa, N.~Ishibashi, H.~Kawai, T.~Kuroki, Y.~Matsuo and T.~Tada,
``Loops versus matrices: The Nonperturbative aspects of noncritical string,''
Prog. Theor. Phys. \textbf{112}, 131-181 (2004)
%doi:10.1143/PTP.112.131
[arXiv:hep-th/0405076 [hep-th]].
%66 citations counted in INSPIRE as of 06 Feb 2023

\end{thebibliography}

%%%%%%%%%%%%%%%%%%%%%%%%%%%%%%%%%%%

\end{document}